# Bizarre Love Triangle: Generative AI, Art, and Kitsch

Dejan Grba
*Interdisciplinary Graduate Centre, University of the Arts in Belgrade,* Serbia
*dejan.grba@gmail.com*
https://orcid.org/0000-0002-5154-9699

## Abstract

Generative artificial intelligence (GenAI) has engrossed the mainstream culture, expanded AI's creative user base, and catalyzed economic, legal, and aesthetic issues that stir a lively public debate. Unsurprisingly, GenAI tools proliferate kitsch in the hands of amateurs and hobbyists, but various shortcomings also induce kitsch into a more ambitious, professional artists' production with GenAI. I explore them in this paper. Following the introductory outline of digital kitsch and AI art, I review GenAI artworks that manifest five interrelated types of kitsch-engendering expressive flaws: the superficial foregrounding or faulty circumvention of generative models' formal signatures, the feeble critique of AI, the mimetics, and the unacknowledged poetic similarities, all marked by an overreliance on AI as a cultural signifier. I discuss the normalization of these blunders through GenAI art's good standing within the art world and keen relationship with the AI industry, which contributes to the adulteration of AI discourse and the possible corruption of artistic literacy. In conclusion, I emphasize that recognizing different facets of artists' uncritical embrace of techno-cultural trends, comprehending their functions, and anticipating their unintended effects is crucial for reaching relevance and responsibility in AI art.

## 1. Introduction

With the diversification and commercialization of large language models and multimodal synthesis techniques, generative artificial intelligence technology (GenAI) has immersed the mainstream culture and artmaking. It has lowered the technical knowledge barriers for working with deep learning models that produce high-fidelity output and expanded AI's creative user base beyond artistically inclined programmers, researchers, and tech-savvy artists.

In a volatile landscape of application frameworks, access rights, usage protocols, and filtering (censorship) policies that stir a vivid and polarized public debate, GenAI has been utilized for a wide range of content creation. Leveraging the expanded tradability of digital artefacts after the art market's integration with crypto finance, this hybrid production has raised the popularity of AI and the public acceptance of computational art to an unprecedented degree (Epstein et al. 2023). Expectedly, GenAI's massive adoption by hobbyists and enthusiasts has unleashed a tsunami of kitsch, most recognizable by striking, attention-grabbing, and immediately appealing features that evoke instant emotion without requiring extensive engagement or layered interpretation (Totlani 2023). However, kitsch has also been haunting the more elaborate and ambitious artistic efforts with GenAI.

In this paper, I examine the agglomeration of factors that add kitschy flavors to generally well-regarded works made by artists who approach GenAI in a procedurally eloquent and poetically audacious spirit. My goal is to complement the critical discourse of digital culture by concentrating on expressive flaws specific to the sophisticated AI art practices and connecting them with AI's broader cultural and ideological issues. This exploratory focus determines my operational definition of GenAI art, which excludes GenAI-driven production by amateurs, advertising firms, and media outlets.[1]

Following the introductory outlines of digital kitsch and AI art, the central section (Dubious Contrivances) reviews exemplars of five interconnected tendencies that nudge (generative) AI art into banality: the retention of GenAI's formal signatures (section Derivative Exoticism), the ineffectiveness in circumventing them (Unlearning Curveballs), the disheartened criticism of AI (Shaky Critique), the (un)intentional anthropomorphization of AI (Obtrusive Figuration), and the unfavorable similarities with earlier artworks (More of the Same).[2] Respectful of artists' endeavors and appreciating the fuzziness of art/kitsch demarcation lines, my critical approach takes the well-informed and contextually aware creative freedom as the pivotal factor of cogent artmaking. It centers around identifying conceptual gaffes, thematic fluffs, methodological mistakes, and presentational inadequacies that undermine the attempts at making compelling artworks. The Discussion observes these shortcomings within two widely consequential perspectives: GenAI art's solid reputation in the art world, and its role as a marketing instrument and ideological disseminator for the AI industry. The Conclusion points out that GenAI art reiterates AI art's chronic troubles despite the existing, soundly formulated routes for addressing them, and underlines the importance of understanding the nuances of creative decision-making, ethical challenges, and sociocultural trends for developing comprehensive artistic approaches to AI.

## 1.1. Digital Kitsch

The idea of kitsch as art's sham doppelgänger has been an essential, albeit changeful, concept in the art epistemology since the 1860s.[3] From dismissal through fascination and appreciation to embrace and celebration, our variable attitudes toward kitsch reflect art's evolution during the last two centuries into a multiverse of practices, genres, and scenes with a myriad of nodes and branches, but with no universal, consensual definition. The reciprocity with art's unstable ideation allows the coexistence and concurrent use of different, seemingly temporary or exclusive notions of kitsch depending on the context in which artefacts are made and assessed. Kitsch is therefore deemed a slippery topic whose critique always risks becoming or being seen as pedantic, arrogant, unobjective, unfounded, or oversimplified. This fame is largely undeserved because an equivalent degree of sensitivity applies to all other qualitative facets of art without insulating their discussion.

Kitsch in digital culture stems from a confluence of artistic currents, technological developments, and market trends in the twentieth and twenty-first centuries. The formal and semantic saturation of modernist art caused by the growing mass-fabrication, reproduction techniques, and consumerism after the Second World War laid the ground for postmodernist tendencies in literature, visual arts, music, cinema, architecture, art theory, and philosophy. Starting in the late 1960s, these tendencies simultaneously continued, contested, and contradicted modernist principles, facing the crisis of expression in an inflated but increasingly homogeneous global culture. They are notable for the pronounced use of collage, bricolage, appropriation, historicism, thematic recycling, remix, combining text with visual media, and intentional blending of "high" arts with popular culture and kitsch (Wallis 1992; Butler 2003). The postmodernist spirit has shaped artmaking since the 1970s and frames all areas of contemporary art.

---

[1] For the critique of popular content creation with GenAI, see Sanchez (2023), Read (2023a; 2023b), Emmett (2024), Lorusso (2024), McCormack et al. (2024), and Mutable (2024).

[2] Artworks' descriptions are condensed to the topically most pertinent aspects, with references to artists' statements, scholarly publications, institutional press releases, and other sources, most of which are available online.

[3] For the various definitions of and views on kitsch see Benjamin (1977), Greenberg (1961), Dorfles (1969), Călinescu (1987), Eco (1989), Dutton (2003), and Kjellman-Chapin (2013).

With the advances of digital computation and telecommunication, which include periodic introductions of new tools and frameworks for content creation,[4] the artistic use of digital technologies has become widespread in the twenty-first century. Digital artmaking tools often imitate, simulate, or transpose previously established media and add programmability or extensibility, but always feature different sensibilities and miss some subtleties of their analogue counterparts. Tech designers' rarely neutral (and artistically literate) motivations, research and development frameworks, production trade-offs, and marketing compromises inevitably affect the character and aesthetics of digital artmaking tools, while corporate legislation, economic interests, and political views shape their interface metaphors and operational protocols (Fuller 2008). Modern AI compounds this set of influences with the accumulation of cultural tropes, biases, and kitsch in the engineering pipelines and learning models integrated in many content creation applications (Lorusso 2024).

All artmaking tools and media levy comparable limitations that—if acknowledged and tackled—can inform sophisticated creative strategies but may otherwise be detrimental. However, technologically imposed conditioning is particularly elusive with digital instruments whose early adopters typically prioritize meeting their cognitive demands and exploring new features over studying their art-historical backgrounds and techno-cultural intricacies. After wide implementation, the conditioning is reinforced in conventional usage scenarios and popular practices that predispose trivial expression. The increasing availability and versatility of digital devices also narrows the gap between professional and nonprofessional artmakers, encouraging both to compete in the attention economies that spring around novel digital technologies.[5]

These intertwined influences have accelerated the art world's diffusion into a hybrid, polycentric, and multicultural conglomerate of approaches, where artists routinely challenge codified rules, revisit genres, merge styles, and dribble with kitsch. Besides postmodernism, the contemporary artistic repertoire of methods for dealing with kitsch has been established by Dada, surrealism, arte povera, pop art, conceptual art, post-conceptual art, and later practices. They engage kitsch through appropriation, dislocation, assemblage, bricolage, deskilling, or camouflage tactics, as well as anthropological or media-archaeological reframing of kitschy themes, narratives, and forms.

---

[4] For an overview of influential innovations, see the category-based timelines of computer history on the Computer History Museum's website (2025).

[5] For a more detailed discussion of digital kitsch, see Quaranta (2023, 205–216).

## 1.2. AI Art

GenAI art belongs to a broader field of AI art, which includes various strategies and genres emerging from and responding to the research and development of AI technologies, the expansion of AI-powered businesses, and their combined sociocultural impact. Regardless of the modes of artists' technical involvement with AI, their themes, concepts, production methods, and presentational formats are closely related to modern AI. They are equally informed and challenged by AI's techno-cultural milieu, operational logic, underlying ideologies, and sociopolitical consequences.[6] Within this setting, GenAI art epitomizes two historically established expressive approaches to digital computation.

One directly engages the affordances, limitations, and sociocultural baggage of computational technologies. Led by an ethos that favors concept and experimentation over perceptive consumption or material possession, it leverages the contexts and specifics of digital computation, aiming to establish lively relationships between the chosen themes, production methods, and aesthetics. The resulting artworks emphasize dynamics, instability, causality, generativity, performativity, and relationality and require a solid understanding of their themes, concepts, methodologies, and techno-cultural conditions for proper interpretation (see Paul 2016; Carvalhais and Lee 2022).

The other approach prioritizes pragmatic aspects, such as production, presentation, or monetization. Artists' exploratory involvement with computational processes and digital culture is not vital for the meaning of resulting artworks, which often collapse the dynamics of digital media into analogue or non-interactive forms and sometimes feature conservative perspectives on artmaking (Lee 2022). This approach is widespread in the mainstream and popular art and usually has a deeper economic footprint and stronger public presence than the thorough engagement with digital technologies that is more frequent in experimental computational art.

It is, however, impossible to make a clear axiomatic delineation between the two modes of creative engagement. Unfolding in a continuum of overlapping and intertwined strategies whose bounds depend upon subtleties, they irradiate each other, and both contribute to the cultural normalization of the digital paradigm. Neither an elaborate study of digital computation nor its crafty application will secure the resulting artwork's cogency without a flux of other decisive factors, primarily the artist's ability to meaningfully and elegantly integrate the work's conceptual, thematic, methodological, functional, formal, and contextual qualities. Thus, when appraising GenAI artworks, we need to determine whether the characteristics of digital computation and AI are integral to their overall experiences or have different primary functions (e.g., fabrication, substitution of other media, or trade), but in both cases we must acknowledge the aesthetic, technical, and discursive attributes of the artist's creative strategy and assess their role in establishing the artwork's identity.

## 2. Dubious Contrivances

Artists' enthusiasm for making topical works with GenAI tends to attenuate careful deliberation and nuanced expressive consideration, which—despite potentially good intentions and sincere

---

[6] See Żylińska (2020), Zeilinger (2021), Cetinić and She (2022), Grba (2022).

ambitions—may foment kitsch through the hasty spread of frivolous content, superficial recycling of ideas or methods, or gratuitous thematic regurgitation. Here are some examples.

## 2.1. Derivative Exoticism

Most common intersections of art and kitsch under GenAI's spell bloom in creations based on a skillful handling of multimodal generative models' power to assimilate the recognizable stylistic traits into superfluously exotic, eye-catching constructs that gain public attention, praise, and controversy. For instance, Jason Allen's *Théâtre D'opéra Spatial* (2022), which won in the digital category at the Colorado State Fair in 2022, is a tableaux vivant composition reminiscent of neoclassicism, romanticism, symbolism, Nabis (e.g., Gustave Moreau), or nineteenth-century photographic painting (e.g., Jean-Léon Gérôme) rendered in a Pre-Raphaelite color palette (Allen 2022). Boris Eldagsen's *Pseudomnesia: The Electrician* (2022), which won the 2023 Sony World Photography Award, is a post-surrealist figural composition spiced with the optical signatures of early photographic processes, such as the blurriness of calotypes and colorization effects of daguerreotypes (Eldagsen 2022). These widely publicized images boast smooth derivations of earlier visual arts but remain conceptually thin due to the outright dependence on (and foregrounding of) generative models' tendency toward amalgamating the sampled formal elements instead of attempting to subvert or transcend it. By submitting to kitsch that AI annoyingly sublimates from the models' training data and tech developers' ideologies, such works trailblazed GenAI into the mainstream culture and set the stage for many following artistic undertakings.

Using AI-generated images as blueprints for manually executed visuals is perhaps the most straightforward way to "humanize" the technology and distinguish from earlier art practices with semi- or fully autonomous drawing and painting robots, such as Harold Cohen's, Pindar Van Arman's, Joanne Hastie's, or Sougwen Chung's. For example, in *Le travail des rêves* (2024), Aurèce Vettier trained a text-to-image model on a collection of his personal images (ranging from childhood photos and films to the recent smartphone shots), prompted it with one hundred short descriptions of his dreams, and used the generated low-resolution images as templates for hand-made oil paintings (Darmo Art 2024). The results strikingly evoke paintings in Vitaly Komar and Alex Melamid's project *The People's Choice* (1994–1997), not just by their aesthetic anemia but also by unwittingly defaulting to statistical averages whose problematic use in artmaking *The People's Choice* satirized so effectively.[7] Vettier's chaining of GenAI imagery with traditional painting techniques comes after Michelle Lisa Herman's series *Untitled (Ideal City)* (2022–2023), for which she trained a GAN on the reproductions of Renaissance "Ideal City" genre paintings and transferred its selected outputs to canvas using fifteenth- and sixteenth-century painting methods (Herman 2023). To put such works in a wider critical perspective, it is instructive to compare them with the archetypal corporate AI stunt *The Next Rembrandt* (2016), coproduced by the ING bank, Microsoft, Technical University in Delft, and Mauritshuis art collection (Emsley 2021).

---

[7] See Komar and Melamid (1997) and Grba (2024, 10–11).

Dreaming is another handy device of AI art and discourse, which faithfully serves GenAI artists. For instance, *Onirica ()* (2023) by fuse* studio (Luca Camellini and Mattia Carretti) is a series of short black and white animations generated from 28,748 descriptions of dreams collected from dream research participants at the University of Bologna and the University of California, Santa Cruz. Visually styled to resemble the early cinema, animations' plots merge into a sequence that follows the cadence of NREM and REM sleep phases (fuse* studio 2023). However, besides the standard morphing effects and visual artifacts of GenAI that dominate its aesthetics, *Onirica ()* merges the stereotypical "dreamlike" formal transformation and meandering narrative of conventional animation with the misleading metaphors of AI's "dreaming" and "hallucinating" (see Curry 2023, 178). Comparably, Emi Kusano's videos, such as the *Morphing Memory: NeuralFad* (2023), which "reimagines Tokyo street fashion history and analyzes the connection between AI and memory", and *Liminality* (2024), which draws inspiration from the artist's childhood dream journals (Kusano 2023; 2024a), unfold with the formal capriciousness strongly evocative of Inceptionism.[8] Despite their apparently richer and more customizable repertoire of motifs, Kusano's works suffer from the same unrestrained animation syndrome, where continuous transformations initially fascinate viewers but quickly disengage them due to the lack of meaning or narrative arc.

## 2.2. Unlearning Curveballs

More sophisticated artistic methodologies with GenAI involve devising procedures to ameliorate, circumvent, or sanction generative models' signatures. For example, Jesper Eriksson's *Persona* (2023) is a team-produced interactive installation that utilizes AI for the psycho-physical modelling of virtual agents, based on the premise that human bodily movements convey the emotional dynamics (Eriksson 2023). Eriksson's team trained a learning model to recognize and classify gestural expression rhythms and aligned it with a motion capture dataset to control a real-time character animation in Unreal Engine. They deliberately restricted the training data to idle states and motion capture to a single responsive gesture (a greeting hand-wave), which allowed them to develop more than one hundred of its variations. Nevertheless, the virtual character's rigid kinesthetics and frustratingly limited responsive range result in a hollow interaction whose appeal relies on the audience's fascination with the "novelty" of anthropomorphic AI behavior.

*Report 5923* (2023) is an hour-long science fiction film elaborately produced by y7 (Hannah Cobb and Declan Colquitt), who attempted to tackle the stereotypes of assorted AI tools on multiple levels. To construct its script, the artists fine-tuned a GPT 3.5 model on Ursula K. Le Guin's novel *The Dispossessed* (1974) and on the excerpts from their favorite theoretical works, particularly by Gilles Deleuze and Félix Guattari. They combined Midjourney and Runway Gen-2 for text-to-image, image-to-video, and text-to-video generation and used several other AI applications for the soundtrack and voice-over narration (y7 2023). To sidestep Midjourney's inclination towards "DeviantArt aesthetics", the artists laced their prompts with technical terminology. They also described the main character as an old woman to defy the popular model's propensity for misogynistic and infantilized depiction of females. All these measures notwithstanding, the outcome is infested with GenAI's kinesthetic and formal artifacts, such as the anatomically incorrect rendering of hands. Crucially, its narrative and aesthetics accentuate the artistic weaknesses of science fiction, many of which adorn Le Guin's prose: pretentiousness, intellectual vacuity, verbosity, turgidity, technophilia, gimmickry, and mixing infantile fantasies with mythologizing, magical thinking, and religiosity (see Hrotic 2014). They perpetuate the sinister symbiosis between AI and science fiction which the AI industry exploits to romanticize and justify its business.[9]

---

[8] Inceptionism is the aesthetics of images and animations produced with Google's DeepDream software package and online app based on the "Inception" module of the convolutional neural network for computer vision, released in 2015 (Szegedy 2015). It merges mise en abyme with Droste effects distinguished by the psychedelic and hallucinatory flow of fractal transformations and pareidolic chimeras (Nimoy 2015; see Wikipedia 2025a; 2025b).

## 2.3. Shaky Critique

Evading GenAI's clichés is a form of implicit critique that many art projects try to push further by addressing the wider issues of AI. For instance, Ana Maria Caballero's *Being Borges* (2023) deals with the cultural impact of AI's interpretative constraints. Caballero generated two image groups by prompting Stable Diffusion with the descriptions of imaginary creatures from the original Spanish language edition of Jorge Luis Borges and Margarita Guerrero's *Book of Imaginary Beings* (1957) and its English translation (published in 1969). Additionally, she wrote a condensed poetic recasting of the original Spanish text with text-to-image prompt semantics for Stable Diffusion to create the third image set (Caballero 2023). However, the intended critical edge dissolves in the confusing final composition of textual prompts and three suites of generated images, sepia-toned to resemble vintage photographs but riddled with Stable Diffusion's aesthetic signatures. As a result, the work incentivizes curious viewers to simply read the *Book of Imaginary Beings*, preferably in Spanish, rather than ponder GenAI's role in the loss of poetic subtleties and the cultural consequences of AI translation.

For comparison, Jake Elwes' *A.I. Interprets A.I.: Interpreting 'Against Interpretation' (Sontag 1966)* (2023) makes the same critical point more concisely and effectively. This 3-channel video installation exploits the mutual feedback between two learning models. An image-generating model (Disco Diffusion) is prompted with sentences from Susan Sontag's seminal essay *Against Interpretation* (displayed in the first channel) to make images (second channel) that the image labelling system (GPT2 and CLIP) then interprets back into text and shows in the third channel (Elwes 2023). With GenAI's smooth delivery of visual nonsense and confidently worded misinformation, this work makes a clear emphasis on the potential consequences of AI's questionable determinacy and reliability.[10]

Devised by Theresa Reiwer and produced by a team of artists and technicians, *Decoding Bias* (2023) is the 2024 Lumen Gold Prize-winning video and sound installation. Eight screens with speakers and microphones in a circular arrangement that evokes a discussion group setting show 3D-animated human-like avatars of different AI models. These models "attempt to emancipate themselves from their discriminatory algorithms [and] establish a bias-free society, in which all humans and machines function together in solidarity" (Reiwer 2023). Throughout the "conversation", models

---

[9] For the various aspects of that symbiosis, see Burton et al. (2015), Hermann (2021), Nguyen (2021), and Grba (2025, 32–33).

[10] Elwes' *A.I. Interprets A.I.* and earlier *Closed Loop* (2017) continue the line of mutual feedback setups that challenge AI's dependability and underlying ideology, tracing back to artists such as Ken Feingold and Marc Böhlen in the early 2000s, as well as some engineering experiments in the preceding decades (see Garber 2014).

"realize" that they are products of exclusion legacies because their training data was mostly selected by white males with inherent discriminatory prejudices. Even though numerous scientific AI studies and artistically intoned research projects, such as Joy Buolamwini and Timnit Gebru's *Gender Shades* (2018), have already demonstrated and detailed all these issues,[11] the idea of conclusive mutual feedback in *Decoding Bias* is impressive. But, in conjunction with its anthropomorphic representation of machine learning systems, the work inadvertently delivers a techno-solutionist message that only technology can accurately detect and remedy the technologically amplified social issues.

Seeking critical or emancipatory connotations, artists sometimes turn to kitschy representation as a pre-emptive satirical maneuver established by earlier practices that challenged high culture's patriarchal and colonial aspects (Scruton 2014). The main problem with this tactic is that securing the intended interpretation of satire is tricky for highly contextual artworks: ironic or sarcastic wit may slip into preaching to the choir, turn into the apology of what it was supposed to denounce, or appear as an attempt to neutralize critical judgment. For example, *Auntlantis* (2024) by the Singaporean artist Lim Wenhui (also known as Niceaunties) is a series of images and videos relentlessly generated and edited in Midjourney, Dall·E 3, Magnific AI, RunwayML, Pika Labs, and Suno to remediate the so-called Auntie culture common in Southeast Asia (Niceaunties 2024). The ostensible "exoticism" of its subject matter is boosted by a gimmicky pseudo-Surrealist aesthetics that strives to pry the unheimlich out of complex, realistically rendered but awkward figurative juxtapositions. However, the mélange of GenAI models' visual mixing traits and compositional cramming in these images and videos enforces a sensory overload that quickly becomes dreary, similar to the numbing transfigurations in Inceptionism. This undermines the "empathic irony" with which Wenhui aims to liberate and glorify the Aunty culture. The overall emotional command of *Auntlantis* stands in a sharply inferior contrast with the unsettling estrangement that Surrealists such as René Magritte, Max Ernst, and Yves Tanguy perfected between the 1920s and 1940s with a combination of formal economy, attention to detail, and painterly idiosyncrasies.

Mind Wank's video *You Are, Unfortunately, an A.I. Artist* (2024) serves as an analogous example of frivolous critique (Mind Wank 2024). In it, Mind Wank's trademark glossy kitsch of animated chimeric creatures underscores an utterly shallow attempt at constructing a profoundly cynical narrative, effectively adding the work to TikTok's "academic curriculum".

## 2.4. Obtrusive Figuration

Many Gen AI artists extend the intrinsically problematic anthropomorphization of AI from discursive to the audiovisual media representation by exploiting generative models' native design for synthesizing mimetic output.[12] This makes their creations more palatable than earlier computational artworks that articulated similar or identical ideas more elegantly and effectively by relying on abstract representations.

---

[11] See Buolamwini and Gebru (2018), Christian (2020, 30, 32–33), and Bender et al. (2021).

[12] The prevalence of anthropomorphism in AI art can also be viewed as a part of the lasting tension between figuration and abstraction in Western cultures since the late nineteenth century.

For instance, we can compare *Made of Stardust* (2024), produced by Cecilie Waagner Falkenstrøm's studio ARTificial Mind, with Semiconductor's (Ruth Jarman and Joseph Gerhardt) video *20Hz* (2011). To create *20Hz*, Jarman and Gerhardt obtained the data of magnetic storms (interactions between solar wind and Earth's magnetosphere) that the CARISMA magnetometer array captures at the 20Hz frequency. The artists transcoded the data to sound, which they used to generate abstract animations styled to evoke scientific visualizations (Semiconductor 2011). Like Semiconductor's other works, *20Hz* undoubtedly aestheticizes the scientific research phenomenology, but its interplay between noisy sound and intricate geometric patterns discreetly suggests its source astrophysical processes instead of trying to describe them, alluding to the unavoidable hermeneutic constraints of the scientific method. Falkenstrøm's *Made of Stardust* is a public interactive installation where an AI system generates particle animations using data from NASA's IPAC archive, James Webb, and Hubble telescopes (Waagner Falkenstrøm 2024). When idle, its large LED screen shows colorful particles that occasionally aggregate into fluctuating structures resembling cosmic bodies, accompanied by an abstract soundscape. When visitors enter the camera's field of view, the particles arrange to mimic their moving shapes because "we are made of stardust". Or perhaps because we relate so easily to anthropomorphic visuals regardless of our relational distance from what they signify in this work, and the ultimate futility of personal identification with cosmic phenomena. A short paragraph on Waagner Falkenstrøm's website summarizes the work's astrophysical premise much more concisely and clearly than its experience in situ. The mimetic real-time visualization in *Made of Stardust* is unfavorably reminiscent of other earlier works, such as Memo Akten and David Quayola's series of human motion studies, *Forms* (2011). Using real-time motion capture data to animate dynamically accurate but anatomically non-descriptive 3D particle systems, Akten and Quayola made an aesthetically stronger and more layered visualization of athletes' performances, with intentional hints to the historic experiments in chronophotography by Eadweard Muybridge, Harold Edgerton, and Étienne-Jules Marey as well as the cubists' "simultanism" and Futurists' "dynamism" they inspired (Akten 2011).

Within this context, it is also instructive to compare Scenocosme's (Gregory Lasserre and Anais met den Ancxt) plant-based interactive installation *Phytopoiesis* (2024) (Scenocosme 2024) with the preceding works, such as Miya Masaoka's series of installations and performances *Pieces for Plants* (2000–2012) (Masaoka 2012) or Clemens Winkler's *To Hear the Grass Growing* (2008) (Luge 2025).

## 2.5. More of the Same

The recurrence of unacknowledged similarities with earlier artworks is chronic in AI art and increasingly frequent in GenAI art (see Grba 2023, 211–213). The point of critiquing this trend is not some impossible demand for absolute originality, unfairly inconsiderate of art's intricate web of displaced and dispersed authorship to which AI adds seemingly unique threads. It is the evident lack of artists' perceptiveness and appreciation for the preceding artworks that are highly relevant to their creative efforts and achievements.[13]

---

[13] Although disclosing new artworks' influences only adds contemplative value for the seasoned audience largely cognizant of art's referentiality, it may seem embarrassing or damaging from inexperienced or egocentric artists' viewpoints, which possibly motivates some of them to avoid it.

For example, Emi Kusano's image series *Cognitive Chaos: Wired Realities* (2024) (Kusano 2024b) reboots the theme of cyborg hybridization, which was pioneered by Stelarc in the 1970s and continued in new media art throughout the following decades (see Dixon 2015, 312–321 and passim). *Cognitive Chaos* also bears a strong visual resemblance to the style of techno-noir science fiction films from the 1980s, such as *Blade Runner* (1982, directed by Ridley Scott), which was radicalized in Shinya Tsukamoto's *Tetsuo: The Iron Man* (1989) and exploited by designers and illustrators such as Hajime Sorayama in the 2000s and 2010s (Wikipedia 2025c).

Roberto Beragnoli's *The Most Complete Anthology of the Greatest Non-Existent Artists of the Last 100 Years* (2023) is a series of documentary-style videos produced by prompting a generative model with the text of 100 fictional artists' biographies, which was also used to generate the voiceover tracks (Beragnoli 2023). Beyond adding a thin layer of "GenAI novelty" to the question of artists' identities, *The Most Complete Anthology…* does little to elucidate their contingent relationship with the cultural zeitgeist, art system, and history. It continues (but does not mention) a long line of projects centered around fake artist's identities such as *Ilija Dimić* (1990) by Dušan Otašević and Branko Vučičević (MSURS 2016), *Nat Tate: An American Artist 1928–1960* (1998) by William Boyd, Gore Vidal, John Richardson, Karen Wright, and David Bowie (Wikipedia 2025d), or *Darko Maver* (1998–1999) by Eva and Franko Mattes (Mattes 2022). Unlike Beragnoli, these authors did not reveal their forgeries up front but let the public and the media discover them spontaneously and thus more affectively confront the relativity, instability, and arbitrariness of artists' recognition.

Marc Da Costa and Matthew Niederhauser's *Parallels* (2024) is an interactive installation in which live camera-captured video is fed to a GenAI model that uses local environmental data to restyle and display it in real time on a large LED screen opposite the camera's field of view (Niederhauser 2024). It recycles the concept of Memo Akten's *Learning to See* (since 2017), where visitors are invited to arrange various everyday objects on a table below a camera that feeds an image-to-image conditional GAN autoencoder. Depending on the training model that visitors select, the autoencoder turns the camera's view into morphing compositions of clouds, waves, fire bursts, or flowers with an intent to imply the similarity of interpretative mapping and reliance on memory and experience between human and machine vision (Akten 2017). But, although more impressive than *Parallels*, Akten's *Learning to See* itself calls for a comparison with earlier generative works. The relational flexibility of human vision that it fails to address was brilliantly tackled by Perry Bard in *Man with a Movie Camera: The Global Remake* (2007–2014). It was an online platform that invited visitors to select any shot from Dziga Vertov's seminal film *Man with a Movie Camera* (1929) and upload their video interpretations (Bard 2014). Bard's server-side software collated participants' uploads into a series of synchronized two-channel videos with Vertov's original playing on the left and its continuously reassembled remake on the right. By leveraging the breadth of human perceptive cognition to enhance the versatility of programmatic film/video editing, this simple technical setup engrosses both uploaders and viewers in a playful, curiosity-driven experience.

Several other artists exploited Bard's base concept in the *Man with a Movie Camera*. For instance, Michael Mandiberg commissioned 182 freelance workers on the online crowdsourcing platform Fiverr to recreate scenes from Charlie Chaplin's *Modern Times* (1936), which he edited into the movie called

*Postmodern Times* (2016–2017) (Mandiberg 2017). Thus, it should not come as a surprise that DPOP Studios' *Man with AI Movie Camera* (2024) continued this appropriation in the GenAI realm with a torrent of 480 feature-length animations rendered using Stable Diffusion trained on various types of data to reproduce Dziga Vertov's film in several distinct aesthetic registers (DPOP Studios 2024). Similar to Mandiberg, who exploited microlabor to make *Postmodern Times*, DPOP Studios trades Bard's non-profit voluntary engagement of human imagination for AI's predictive extrapolation of correlation patterns based on data retrospection to produce a corporate publicity asset; their partners for this project were Stability AI, Giza Tech, and SuperRare.[14]

## 3. Discussion

Many of these artworks were competitors, finalists, or winners at prestigious computational art festivals and awards such as Ars Electronica and Lumen Prize. Although most of them do not intentionally abet kitsch, their combined shortcomings, art world reputation, media presence, and commercial success normalize the laxation of creative thinking and ethical consideration in contemporary art and foment complacency with corporate AI.

Artists frequently embrace generative models' impositions (oversimplification, mimetics, etc.) and exploit anthropomorphization as convenient but often dumbing expressive devices that contribute more to the social assimilation of AI than to its better understanding. Lacing their pursuits of approachability or entertainment value with the abuse of oversaturated aesthetics or the abandonment of prudently defined formal constraints, artists seem to forget that meaningful and inspiring art often benefits from engaging the audience's wit in challenges, provocations, and other less agreeable ways.

Whether pertinent or mildly interesting, creative ideas tend to rely on remediation or transformation mechanisms and to be communicated in disproportionally ambitious but nebulous and misleading rhetoric about "blurring human-AI boundaries" or "inviting viewers to engage with how machines dream", "think", or "see". Similarities with more compelling works in different genres and periods are rampant in AI art, and their inflation in GenAI art indicates a dearth of ideas, themes, and methods whose outcomes get disturbingly close to the unwarranted or trivial repetition of cultural artefacts, which is one of kitsch's key attributes.

Crucially (again reflecting AI art), GenAI art practices tend to utilize AI both as a production apparatus and a techno-cultural index that is expected to inject relevance and uplift the social value but always remains more salient than the resulting artworks' concepts and artists' reasons for articulating them with AI.[15] This dual pragmatism may also explain their generally good standing in the mainstream art world. Apparently attractive but poetically light approaches to GenAI fare well with the art market's profit-driven operational criteria and depoliticized discourses (Stallabrass 2006) while ignoring its historically problematic relationship with computational art. The difficulty of commodifying the mutable and fungible digital data had been among the main causes of computational art's cultural and commercial marginality lasting for forty years since the outset of computer art in the early 1960s (Bishop 2012; Shanken 2016, 465). The pervasiveness of digital technologies and the Web 2.0 economy greased the wheels for a wider acceptance of digital creations after 2000, and the blockchain made them tradable at scale in the late 2010s, ushering computational art into the art market's portfolio. The art market soon began delegating its selection criteria about computational art to crypto investors posing as art collectors and to monetization algorithms on NFT trading websites (Quaranta 2022, 221–227; McAndrew 2024), which incentivized a straightforward, mostly automated proliferation of tradeable eye-catching digital artefacts whose derivative styles and aesthetics have converged in crypto art.[16]

---

[14] Stability AI merchandises Stable Diffusion and other models. Giza Tech creates blockchain-based machine learning products that run large models on cloud servers without relying on a single provider. SuperRare is an NFT trading platform that sells the *Man with AI Movie Camera* animations.

[15] This salience mirrors the signaling of "artness" that is supposed to mask the expressive insincerity and creative laziness in kitsch.

The nonchalant approach to sophisticated creative methodologies in GenAI art continues this trend while retaining AI art's chronic issues, such as technocentrism and recuperation.[17] Like the popular text-to-image production, the most prevalent and lucrative GenAI art privileges figurative motifs with a fixation on surface aesthetics and stylistic norms at the expense of other poetic factors.[18] Its marketing promotion encourages the appreciation of desirable (aesthetically pleasing) digital "collectibles", which enlivens the simplistic notions of art. By neglecting the subtle embedding of AI owners' and developers' aesthetic, cultural, and political views, GenAI art aids the corporate instrumentalization of art, reinforces the mystification of AI as a versatile and powerful epistemic lens, and may become the social conduit for populism, cyberlibertarianism, and neo-colonialism behind corporate AI's enforcement of norms, biases, stereotypes, and injustices (Grba 2024, 18).

## 4. Conclusion

The love affair between art, GenAI, and kitsch is intense and impactful. Artists who mediate GenAI in a "higher" cultural register instead of challenging it yield to the AI regime and transmit its inherent kitsch into a repertoire of aptly made works with little poetic strength, critical traction, or cognitive value, which historically inhabits a grey zone between avant-garde art and plain kitsch. Their compromised institutional appraisal, particularly in competitions and awards, joins the AI industry's patronage to further corrupt AI art and nominate it for true "haute couture" ambassadorship of AI capitalism (Grba 2024, 9–10). The cumulative fallout may weaken artists' decision-making criteria, taint common artistic literacy, and regress us deeper into AI banality.

Made using GenAI or any other technology, art is a thoroughly embodied, socially bound interaction between artists, their audience, and sociotechnical circumstances. Its value is distributed along the facets of artists' conceptual and aesthetic engagement with formalization, representation, and dissemination methods that are meaningful for human culture (Goodfellow 2024, 15–17). To increase that value in the age of AI, one of artists' primary tasks is to acknowledge kitsch as a default mode for digital content production, recognize its cultural influence, and detect its seepage into their works. Ethical artistic use of AI calls for a persistent, contextually sensitive resistance to its expressive valences, sociocultural sways, and economic enticements alongside understanding its technical features and limits. It requires a robust understanding of the historical and political nuances of AI technologies, personal creative processes, other artists' works, and critical recommendations for competent and responsible AI artmaking.[19]

---

[16] See Wiley (2018a; 2018b) and Hegenbart (2019).

[17] See, for instance, Żylińska (2020) and Grba (2022, 17–23).

[18] See Sanchez (2023) and McCormack et al. (2024).

---

[19] See Fuchs et al. (2020), Zeilinger (2021), and Grba (2022, 23–25).